\documentclass[pra,balancelastpage,twocolumn,footinbib,superscriptaddress]{revtex4-1}
\usepackage{color,amsthm,amsmath,amsfonts,graphicx,bm}
\usepackage{bm}
\usepackage{color}
\usepackage{amsfonts}
\usepackage{amssymb}
\usepackage{amsmath}
\usepackage{epsfig}
\usepackage{graphicx}
\usepackage{enumerate}
\usepackage{multirow}
\usepackage{verbatim}
\usepackage{color}
\usepackage{subfigure}

\theoremstyle{plain}
\theoremstyle{definition}
\begin{document}

\title{Exploring Thouless Pumping in the Generalized Creutz Model: A Graphical Method and Modulation Schemes}

\author{Yan-Jue Lv}
\affiliation{College of Physics and Electronic Engineering, Institute of Solid State Physics, Sichuan Normal University, Chengdu 610068, China}
\author{Yang Peng}
\affiliation{College of Physics and Electronic Engineering, Institute of Solid State Physics, Sichuan Normal University, Chengdu 610068, China}
\author{Yong-Kai Liu}
\affiliation{College of Physics and Information Engineering, Shanxi Normal University, Taiyuan 030031, China}
\author{Yi Zheng}
\email{Corresponding author: zhengyireal@sicnu.edu.cn}
\affiliation{College of Physics and Electronic Engineering, Institute of Solid State Physics,  Sichuan Normal University, Chengdu 610068, China}

\begin{abstract}
Thouless pumping with nontrivial topological phases provides a powerful means for the manipulation of matter waves in one-dimensional lattice systems. The band topology is revealed by the quantization of pumped charge. In the context of Thouless pumping, we present a graphical representation for the topological phases characterized by the Chern number of an effective two-dimensional band. We illustrate how the two topological phases with distinct Zak phase is connected in the pumping process. Such a visual depiction exhibits typical patterns that is directly related to a linking number and to the Chern number, allowing for the construction of Thouless pumping schemes in a practical way. As a demonstration, we present a generalized Creutz model with tunable Peierls phase, inter-leg imbalance and diagonal hopping. Various modulation schemes for Thouless pumping are studied, focusing on their graphical representations in Bloch space, as well as the quantized pumping phenomenon in real space.
\end{abstract}

\maketitle

\section{Introduction}\label{intro}
The concept of topology has been introduced in the study of quantum systems for decades, paving the way for the exploration of exotic phases of matter that challenge conventional understanding \cite{hasan2010colloquium, qi2011topological}. Topological phase transition can occur without breaking any symmetry, which transcends the traditional Ginzburg-Landau paradigm, as exemplified by the quantum Hall effect \cite{thouless1982quantized, niu1985quantized, avron2003topological, kane2005quantum}. The global properties inherent to a topological phase confer remarkable robustness against local perturbations, making these phases particularly intriguing for both fundamental research and practical applications. This robustness has catalyzed the investigation of topological insulators, materials that promise revolutionary advancements in electronic and spintronic devices due to their unique surface states \cite{konig2007quantum, zhang2009topological, hsieh2009observation, sato2017topological}. Various quantum models have been proposed to elucidate the topological characteristics of energy bands. In one-dimensional (1D) systems, the Su-Schrieffer-Heeger (SSH) model stands out as a fundamental framework, exhibiting non-trivial topology through a non-zero quantized Zak phase \cite{su1979solitons, zak1989berry, ryu2002topological, xiao2010berry, atala2013direct}. This non-trivial phase leads to the presence of zero-energy edge modes, which merge into the bulk at the phase transition point, highlighting the intricate interplay between topology and quantum mechanics. In two-dimensional (2D) systems, band topology is typically characterized by the Chern number, which is obtained by integrating the Berry curvature over the Brillouin zone (BZ) and has a direct correlation with the Hall conductivity \cite{thouless1982quantized}. A seminal model that supports 2D topological phases is the Haldane model \cite{haldane1988model}, which has been successfully simulated in cold atom experiments \cite{jotzu2014experimental, cooper2019topological}, demonstrating the feasibility of realizing these exotic states in controlled environments.

Thouless pumping is an intriguing manifestation of non-trivial band topology. It refers to the 1D charge transport phenomenon in a quantized manner as a response to periodic modulation of the quantum system. Thouless pumping has been studied in diverse engineered systems ranging from quantum gases to photonics and electric circuits \cite{thouless1983quantization, niu1984quantised, nakajima2016topological, lohse2016thouless, citro2023thouless, kraus2012topological, imhof2018topolectrical, shah2024colloquium}. However, specific models for such a quantum pump are still rare. A prominent example is based on the Rice-Mele scheme \cite{rice1982elementary, wang2013topological, nakajima2016topological, lohse2016thouless}, which involves periodic modulation of hopping rates and on-site potentials within the framework of the SSH model. By treating time as an additional quasi-momentum axis, one can construct a 2D BZ, where the pumped charge is quantized according to the Chern number, a hallmark of topological order in 2D systems. As the 1D model evolves, the inherent chiral symmetry may be broken, leading to a situation where the Zak phase becomes unquantized. This transition raises important questions about the connection between the Zak phase and the Chern number, as well as the characterization and the robustness of topological features in intermediate models.

On the other hand, various extensions to the basic models have been proposed to study richer topological phases and to explore Thouless pumping schemes. One notable extension of the SSH model is the cross-linked two-chain ladder, commonly referred to as the Creutz model \cite{creutz1999end, asboth2016short}. This model has emerged as a paradigmatic example for investigating topological phenomena due to its unique structure \cite{junemann2017exploring, kuno2020extended, zurita2020topology}. Additional hopping phases and inter-leg detuning can be implemented in such a quasi-1D system, allowing for the investigation of topological phase transition, bulk-edge correspondence and transport properties in the quantum Hall regime. Furthermore, quantized charge pumping through phase modulations has been proposed \cite{sun2017quantum}, offering insights into the fundamental manipulation of matter waves in topological systems. To simulate the Creutz model, ultracold atoms in optical lattices provide a powerful platform with parameters controllable to a great extent \cite{junemann2017exploring, gross2017quantum, schafer2020tools, kang2020creutz}. Notably, artificial gauge fields can be synthesized through laser-assisted tunneling \cite{aidelsburger2011experimental, jimenez2012peierls, miyake2013realizing, aidelsburger2013realization, atala2014observation, galitski2019artificial}, facilitating the exploration of non-trivial topological features and their implications for quantum transport.

In this work, we present a graphical representation of topological characterization, allowing for the definition of a winding number even in the absence of chiral symmetry in 1D systems. This framework provides a connection between the Chern number and a linking number in Bloch space. Based on such an understanding, we show how Thouless pumping scheme can be constructed by taking the generalized Creutz model as a demonstration.

The paper is organized as follows. In Sec. \ref{sec2}, we give a general formalism to describe the topological feature of static and periodic modulated models, emphasizing the graphical representation as a criterion for topological pumping. In Sec. \ref{sec3}, we present a generalized Creutz ladder which supports distinct topological phases. In Sec. \ref{sec4}, three pumping schemes based on such a model are proposed, showing typical winding and linking patterns in Bloch space. Our conclusions are summarized in Sec. \ref{sec5}.
\begin{figure*}[tbp]
	\includegraphics[width=14cm, height=3.9cm]{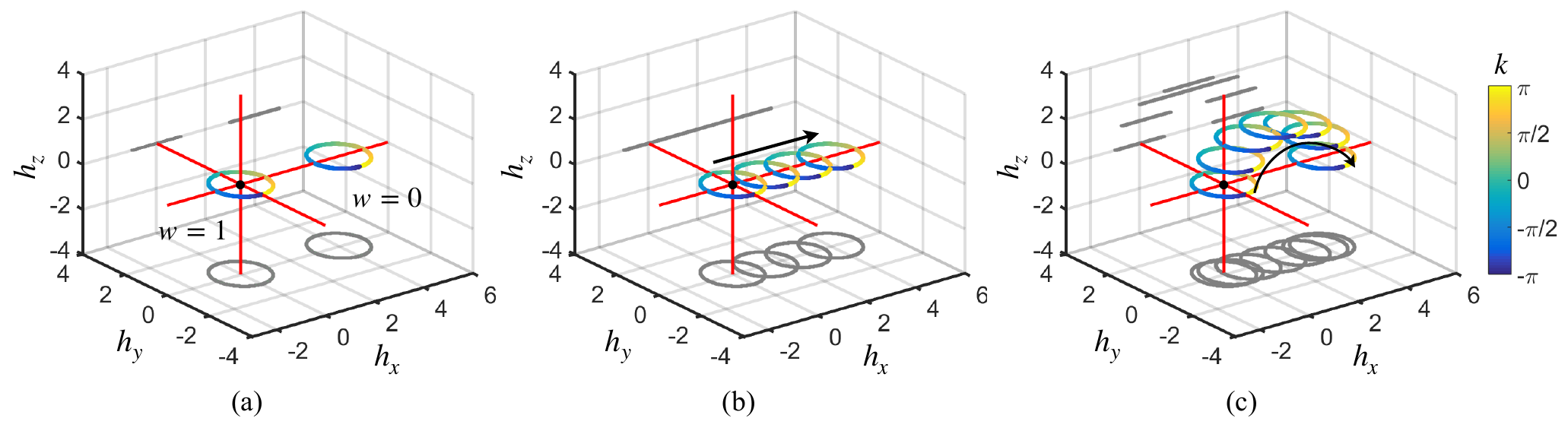}
    \caption{ Schematic demonstration of how the two topological distinct phases are connected by concentrating on $\bm{h}(k)$. The colored circles in all plots represent the endpoints of $\bm{h}(k)$ and the colorbar indicates the value of $k$. The gray lines are projections of the $\bm{h}$-circles. (a) The two topological phases with $w=1$ and $0$. (b) $\bm{h}$-circles lie in the $h_x$-$h_y$ plane (with $h_z=0$) as parameters vary. The process encounters a gap closing point as the $\bm{h}$-circle crosses the origin. (c) The route which breaks chiral symmetry. }
\end{figure*}

\section{General formalism for Topological Phases and Thouless pumping}\label{sec2}
Let us consider a 1D or quasi-1D system consist of two-partite lattice sites. Under periodic boundary condition, the Hamiltonian in quasi-momentum space is given by
\begin{equation}\label{Hamiltonian}
	H(k) = \varepsilon\cdot\bm{1}+\bm{h}(k)\cdot\bm{\sigma}.
\end{equation}
Here $\bm 1$ and $\bm \sigma$ are identity matrix and the three Pauli matrices, respectively.
$k$ represents the quasi-momentum restricted in the first BZ. The band topology of the system is essentially determined by the chiral nature of $\bm{h}(k)$. Specifically, $H(k)$ gives rise to a mapping from $k$-space ($S^1$) to the manifold of Bloch wavefunction. To define a nontrivial topology, $\bm{h}$ has to be a two-dimensional vector. In the SSH model, for example, we have $h_z=0$ as a result of reflection symmetry and time-reversal symmetry of the bulk Hamiltonian \cite{zhai2021ultracold}. For the Creutz model, the $\sigma_y$-term vanishes since the system is invariant under $\mathcal{R}\sigma_x$ and $\mathcal{T}\sigma_x$ \cite{hugel2014chiral, zheng2017chiral}. Here  $\sigma_x$ represents quasi-spin reversal, $\mathcal{R}$ and $\mathcal{T}$ are respectively the operations of mirror reflection and complex conjugation. In these cases, the topological feature is determined by the winding of $\bm h(k)$ in a plane as $k$ varies across the first BZ, which can be characterized by the Zak phase $\varphi_\text{Zak}=i\int_{\text{BZ}}\langle u_k|\partial_k|u_k\rangle dk$, with $|u_k\rangle$ the two-spinor eigenstate of $H(k)$ corresponding to the lower band. At this point, the Zak phase is quantized and so is the winding number ($w\equiv \varphi_\text{Zak}/\pi\in\mathcal{Z}$). As the Hamiltonian (\ref{Hamiltonian}) lacks one component of $\bm \sigma$, the system possesses chiral symmetry, \textit{i.e.,} there  exist an operator $\mathcal{C}$ that anticommutes with the bulk Hamiltonian and satisfies $\mathcal{C}^2=\bm{1}$. The chiral symmetry is associated with the appearance of zero energy edge states according to the bulk-boundary correspondence \cite{ryu2002topological, delplace2011zak}. The symmetry class of the topological insulating phase can be the chiral unitary (AIII) and the BDI class for the Creutz model and the SSH model, respectively \cite{chiu2016classification}.

Suppose that $\bm h(k)$ winds in the $h_x$-$h_y$ plane with $h_z=0$, typically as $k$ varies in $(-\pi,\pi]$, the endpoints of vectors $\bm h(k)$ constitute a closed loop ($\bm{h}$-circle). Figure. 1(a) schematically demonstrates two cases of topological distinct phases as labeled by $w=1$ (non-trivial) and $w=0$ (trivial). The topological invariant $w$ count the number of topological defects (the origin in this case) encircled by the loop. Let us assume that some parameters of $\bm h$ vary with time. In order to connect the two different topological phases, the system has to either encounter a gap closing point or experience chiral symmetry breaking. The two possible routes are shown in Fig. 1(b) and (c). During the latter route with finite value of $h_z$, the Zak phase is no longer quantized and the topological feature is believed to be trivial. However, we can still associate each circle in Fig. 1(c) with a winding number by introducing local coordinate transformations.

Concretely, for each certain $k$, we define $\tilde H(k)=U^\dag H(k)U$ and $|\tilde u_k\rangle=U^\dag |u_k\rangle$ with $U$ an unitary operator expressed by
\begin{equation}
	U=\exp\left[ -i\frac{\beta}{2}\left(\cos\alpha\sigma_x-\sin\alpha\sigma_y\right) \right],
\end{equation}
where $\alpha$ and $\beta$ are given by $\alpha=\arccos (h_y/\sqrt{h_x^2+h_y^2})$ and $ \beta=\arctan (h_z/\sqrt{h_x^2+h_y^2})$. This transformation simply rotates the coordinate axes around $(\cos\alpha \hat e_{x}-\sin\alpha\hat e_{y})$ by an angle $\beta$ in the Bloch space. Consequently, the endpoints of $\bm{\tilde h}(k)\equiv U^\dag\bm h(k)U$ locate in the $\tilde h_x$-$\tilde h_y$ plane with $\tilde h_z=0$ in the new defined coordinates. As the chiral symmetry is recovered in $\tilde H$, we can define a winding number
\begin{equation}
	\tilde w=\frac{i}{\pi}\int_{\text{BZ}}\langle \tilde u_k|\partial_k|\tilde u_k\rangle dk,
\end{equation}
 which is again quantized. We have $\tilde w=1$ for the situation where $\bm{h}$-circle surround the $h_z$ axis. The existence of edge states but with finite eigen-energies is expected in such a case. The boundary between $\tilde w=0$ and $\tilde w=1$ can be given by $h_x=h_y=0$. Besides, the original Zak phase acquires an additional value as $\varphi_\text{Zak}=\tilde w\pi + \eta$ with
\begin{eqnarray}
	\eta &=&i\int_{\text{BZ}}\langle \tilde u_k|(U^\dag\partial_k U)|\tilde u_k\rangle dk\nonumber\\
	&=&i\int_{\text{BZ}}\langle u_k|(\partial_k U)U^\dag|u_k\rangle dk.
\end{eqnarray}

As parameters in $\bm{h}(k)$ vary, $\varphi_\text{Zak}$ changes with time. We consider a periodic modulation of parameters, \textit{i.e.}, $\bm{h}(k,t+T)=\bm{h}(k,t)$ with $T$ the period. The route of $\bm{h}$-circle constitutes a closed path. In a topological Thouless pumping scheme, such a path connect two distinct topological phases. The pumped charge within a period is quantized to the Chern number, which is also associated to the accumulated change of Zak phase. Therefore, we have
\begin{equation}
	C=\frac{1}{2\pi}\int_{0}^{T}dt\frac{\partial }{\partial t}\varphi_\text{Zak}=\frac{1}{2\pi}\int_{0}^{T}dt\frac{\partial }{\partial t}\eta.
\end{equation}
\begin{figure}[tbp]
	\includegraphics[width=8.5cm, height=6.8cm]{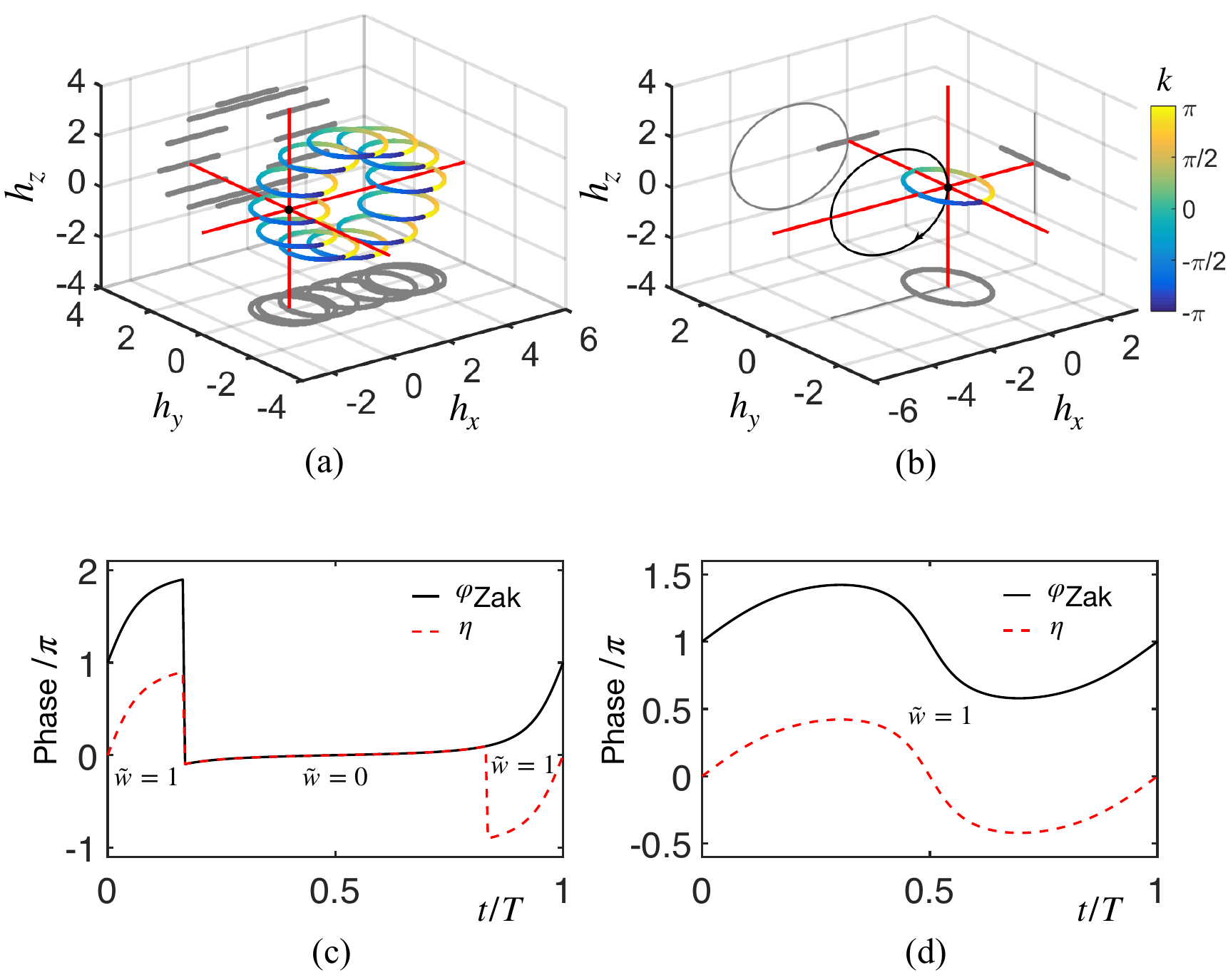}
    \caption{(a) Evolution of the $\bm{h}$-circle for the toy model given in Eq. (\ref{toymodel}). The $\bm{h}$-circle is constituted by the endpoints of $\bm{h}(k,t)$ in Bloch space. (b) Linking between the initial $\bm{h}$-circle and the relative trajectory of the origin. The grey lines are projections. The color bar highlights the value of $k$ in both plots. (c) Zak phase $\varphi_\text{Zak}$ and additional phase $\eta$ with respect to $t$ for the toy model (\ref{toymodel}) with $b=2$ and $a=1$. (d) Same as (c) but with $b=0.4$ and $a=1$. The difference between $\varphi_\text{Zak}$ and $\eta$ yields the extended winding number $\tilde w$. }
\end{figure}

To present an illustrative example, we introduce a toy model with $\bm{h}(k,t)$ given by
\begin{eqnarray}\label{toymodel}
	h_x &=& a\sin k + b(1-\cos(\omega t)),\nonumber\\
	h_y &=& a\cos k,\nonumber\\
	h_z &=& b\sin (\omega t).
\end{eqnarray}
Here $\omega=2\pi/T$ and we have set the energy unit to $1$. The $\bm{h}$-circle is perpendicular to the $h_z$-axis. The trajectory of the center forms a circular loop with a radius $b$. It can be verified that $C=1$ for $b>a/2$. Instead of calculating the Chern number, the variation of $\bm{h}$-circle in a modulation period, as shown in Fig. 2(a), exhibits a paradigmatic example of nontrivial topology. We have set $b=2$ and $a=1$. Alternatively, we can consider the relative position between the $\bm{h}$-circle and the origin, and regard the modulation as the evolution of the origin. As demonstrated in Fig. 2(b), the relative trajectory of the origin form a circle which links with the $\bm{h}$-circle, giving rise to a linking number $1$, coinciding with $C=1$.

In order to show how the two topological different phases with $\varphi_\text{Zak}=0$ and $\pi$ are connected in the pumping scheme, in Fig. 2(c), we plot $\varphi_\text{Zak}$ and $\eta$ with respect to $t$ for the modulation shown in Fig. 2(a). Clearly, the system experiences two transitions between $\tilde w=1$ and $0$. The boundaries at $t/T=\arccos(1-a/b)/(2\pi)$ and $t/T=1-\arccos(1-a/b)/(2\pi)$ are in accordance with the moments when $\bm{h}$-circle intersects the $h_z$-axis. The phase changes of both $\varphi_\text{Zak}$ and $\eta$ accumulate to $2\pi$. In comparison, for $b=0.4$ and $a=1$ ($b<a/2$), there is no transition in $\tilde w$ during a complete cycle. The accumulated phase change is $0$, indicating a trivial case of $C=0$.

We mention that the $\bm{h}$-circles as shown in Fig. 1 and Fig. 2(a) can be translated and reshaped while preserving its topological properties. Such a graphical representation not only provides an intuitive understanding of the topological invariant Chern number, but may also holds promise for exploring additional pumping strategies. In the following, we present three Thouless pumping schemes based on the generalized Creutz model.

\section{generalized Creutz ladder}\label{sec3}
\begin{figure}[tbp]
	\includegraphics[width=8cm, height=2.8cm]{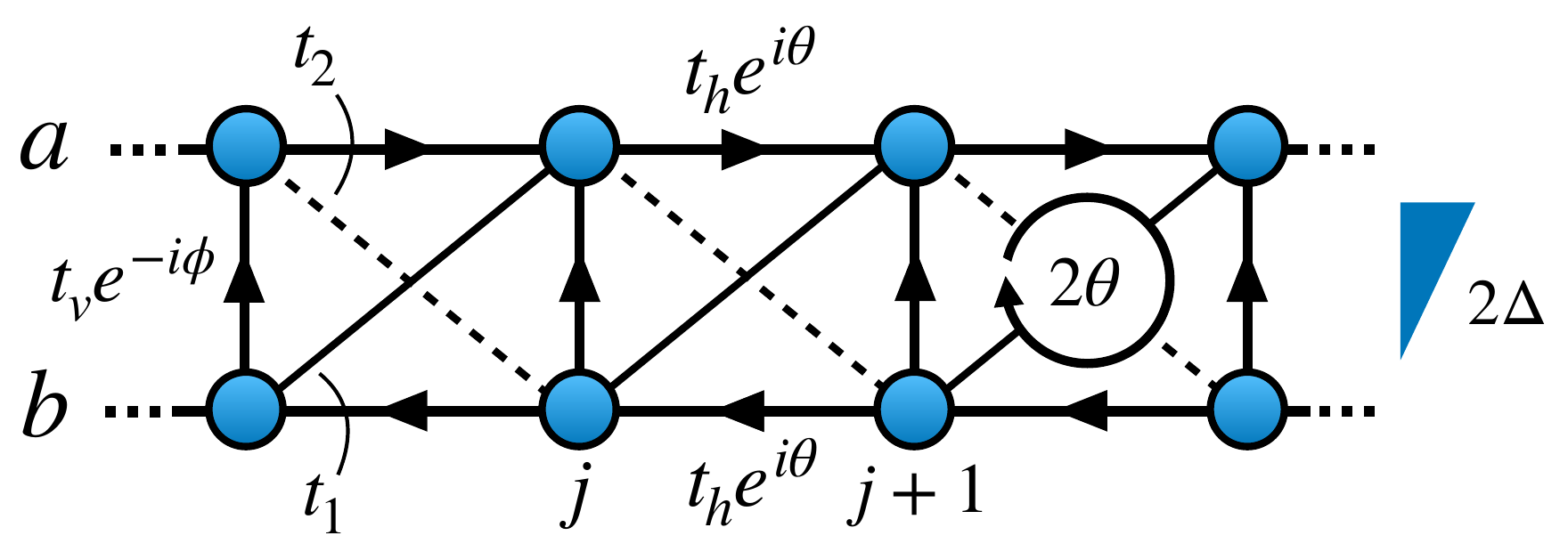}
    \caption{Schematic diagram of the generalized Creutz model with hopping strength, imprinted phases and inter-leg imbalance labeled in accordance with Eq. (\ref{Creutz_model}). }
\end{figure}
The standard Creutz model characterizes a system of spinless fermions on a cross-linked two-leg ladder \cite{creutz1999end}, which was proposed to understand the features of chiral fermions in the context of lattice gauge theories \cite{creutz1994surface, kaplan1992method}. As schematically demonstrated in Fig. 3, we consider a generalized model described by
\begin{eqnarray}\label{Creutz_model}
	\hat H&=&\sum_j \left[t_h(e^{-i\theta}\hat a_{j}^\dag\hat a_{j+1}+e^{i\theta}\hat b_{j}^\dag\hat b_{j+1})+t_ve^{-i\phi}\hat a_j^\dag\hat b_j\right]\nonumber\\
	&+&\sum_j (t_1 \hat b_j^\dag\hat a_{j+1}+t_2\hat a_j^\dag \hat b_{j+1})+\text{H.c.}\nonumber\\
	&+& \sum_j\Delta(\hat a_j^\dag \hat a_j - \hat b_j^\dag\hat b_j),
\end{eqnarray}
where $\hat a_j^\dag$ and $\hat b_j^\dag$ are creation operators for particles on site $j$ of the upper and lower legs. The imprinted phases in horizontal links leads to a net $2\theta$ flux on a square unit cell. In addition, the inter-leg hopping is accompanied by a Peierls phase $\phi$. We have also included an energy imbalance $2\Delta$ between the two legs. In order to realize such a cross-linked ladder system, several experimental schemes have been proposed based on atomic platform \cite{junemann2017exploring, sun2017quantum, kang2020creutz}, photonics \cite{zurita2020topology} and circuit QED \cite{alaeian2019creating, hung2021quantum}.

By taking a Fourier transform, the Hamiltonian can be written as $\hat H=\sum_j\hat\psi_k^\dag \mathcal H_k\hat \psi_k$, where $\hat\psi_k^\dag=(\hat a_k^\dag,\hat b_k^\dag)$ and the Hamiltonian in momentum space is given by $\mathcal H_k=\varepsilon(k)\cdot\bm{1}+\bm{h}(k)\cdot\bm{\sigma}$. We obtain $\varepsilon(k) =2t_h\cos\theta\cos k$ and
\begin{eqnarray}\label{hxyz}
	\nonumber\\
	h_x &=& t_v\cos\phi+(t_1+t_2)\cos k,\nonumber\\
	h_y &=& t_v\sin\phi+(t_1-t_2)\sin k,\\
	h_z &=& 2t_h\sin\theta\sin k + \Delta. \nonumber
\end{eqnarray}
Here we have set the lattice constant $a=1$. A two-band spectrum is given by $E_\pm=\varepsilon(k)\pm |\bm h(k)|$. For $\phi=0$, $t_1=t_2\equiv t_d$ and $\Delta=0$, the Hamiltonian reduces to the conventional Creutz model with $\bm{h}(k)$ winds in the $h_x$-$h_z$ plane. The $\bm{h}$-circle crosses the origin at $t_d=t_v/2$, corresponding to a topological phase transition point. Specifically, we have $\varphi_\text{Zak}=\text{sign}(\theta)\pi$ for $t_d>t_v/2$ and $\varphi_\text{Zak}=0$ for $t_\text{diag}<t_v/2$. The generation is considered aiming to construct Thouless pumping by connecting two of the topological distinct phases. Therefore, we present three schemes based on the combinations of phase modulation ($\phi\neq0$), imbalance control ($\Delta\neq0$) as well as hopping adjustment.

\section{modulation schemes and quantum pump}\label{sec4}
\begin{figure}[tbp]
	\includegraphics[width=8.5cm, height=7.8cm]{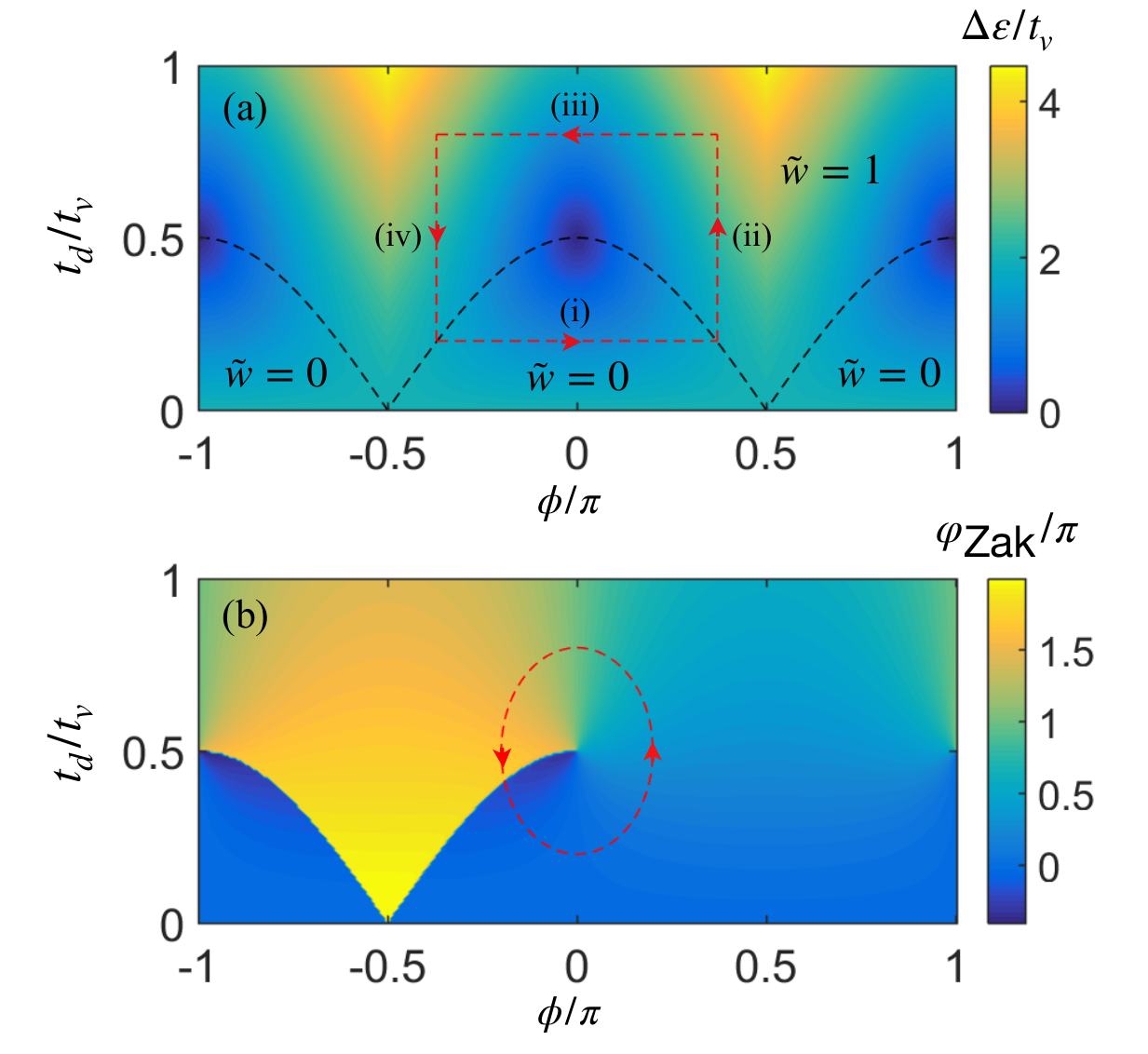}
    \caption{Parameter space spanned by Peierls phase $\phi$ and diagonal hopping strength $t_d$. (a) Value of band gap $\Delta\varepsilon$ with respect to $\phi$ and $t_d$. The black dashed line separates the regimes of $\tilde w=0$ and $1$. The four paths labeled by (i)-(iv) are further demonstrated in Fig. 5. (b) Zak phase $\varphi_\text{Zak}$ as a function of $\phi$ and $t_d$. The closed path represents a typical modulation pattern for Thouless pumping.}
\end{figure}
\subsection{quantum pump by phase modulation and hopping adjustment}
\begin{figure}[tbp]
	\includegraphics[width=8.6cm, height=7.6cm]{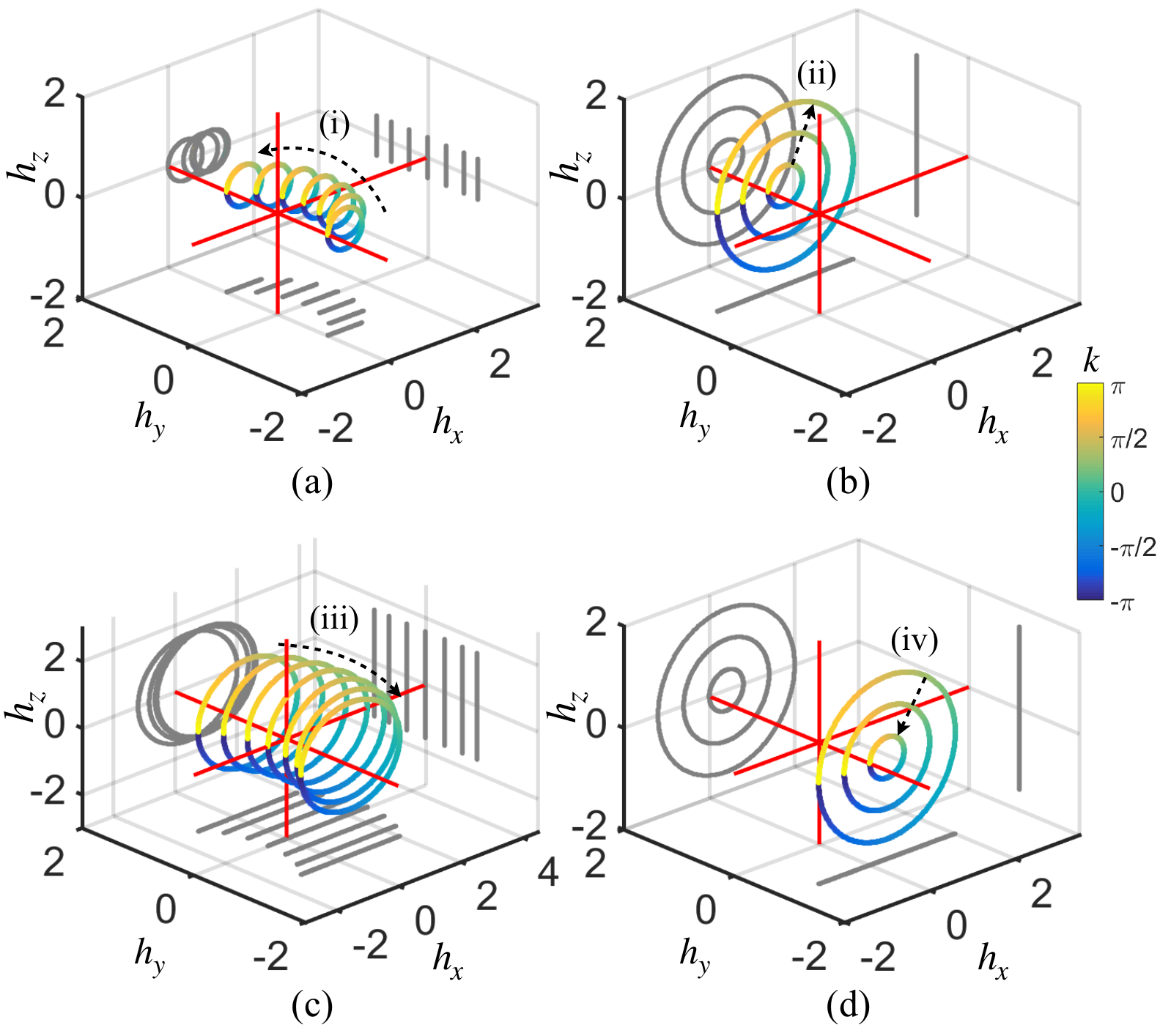}
    \caption{ Variation of the $\bm{h}$-circle following the four paths (i)-(iv) in Fig. 4. The arrows indicate the direction of changes. Projections are plotted on the bottom, rear and right side. Color bar is shared by all plots.}
\end{figure}
We first consider the case of $t_1=t_2=t_d$ and $\Delta=0$. Taking into account the experimental implementation proposed in Ref. \cite{junemann2017exploring}, here we take $t_h=t_d$. The system is governed by
\begin{eqnarray}
	\nonumber\\
	h_x &=& t_v\cos\phi+2t_d\cos k,\nonumber\\
	h_y &=& t_v\sin\phi,\\
	h_z &=& 2t_d\sin\theta\sin k. \nonumber
\end{eqnarray}
The Peierls phase $\phi$ breaks the chiral symmetry. Finite value of $h_y$ shifts the $\bm{h}$-circle out of the $h_x$-$h_z$ plane in a similar manner of Fig. 1(c) [see Fig. 5(a) and (c) for illustrations]. The size of the $\bm{h}$-circle can be changed by adjusting the hopping strength $t_d$. Straightforwardly, we have $\tilde w=1$ for $t_d/t_v>|\cos\phi|/2$. Figure 3 depicts the band gap $\Delta\varepsilon=\min (E_+)-\max(E_-)$ and Zak phase $\varphi_\text{Zak}$ with respect to $\phi$ and $t_d$. We have set $\theta = \pi/2$ without loss of generality. The black dashed line ($t_d/t_v=|\cos\phi|/2$) signifies the boundary between $\tilde w=1$ and $\tilde w=0$. To demonstrate how the $\bm{h}$-circle varies with $\phi$ and $t_d$, we consider a closed path in parameter space as shown in Fig. 4(a). The trajectories of $\bm{h}$-circle corresponding to the four routes (i)-(iv) are respectively exhibited in Fig. 5(a)-(d). In process (i), the $\bm{h}$-circle bypasses the origin since parameters are in $\tilde w=0$ regime. For the moment of $\phi=0$, the topological feature is characterized by $\varphi_\text{Zak}=0$. At the transition points $t_d/t_v=0.2$ and $\phi=\pm 0.369\pi$, the $\bm{h}$-circle touches the $h_y$-axis. In contrast, the $\bm{h}$-circle encompasses the $h_y$-axis in processes (ii)-(iv), in accordance with $\tilde w=1$. Therefore, the $\bm{h}$-circle covers the origin once during a period of the modulation.

\begin{figure}[tbp]
	\includegraphics[width=8.5cm, height=5.7cm]{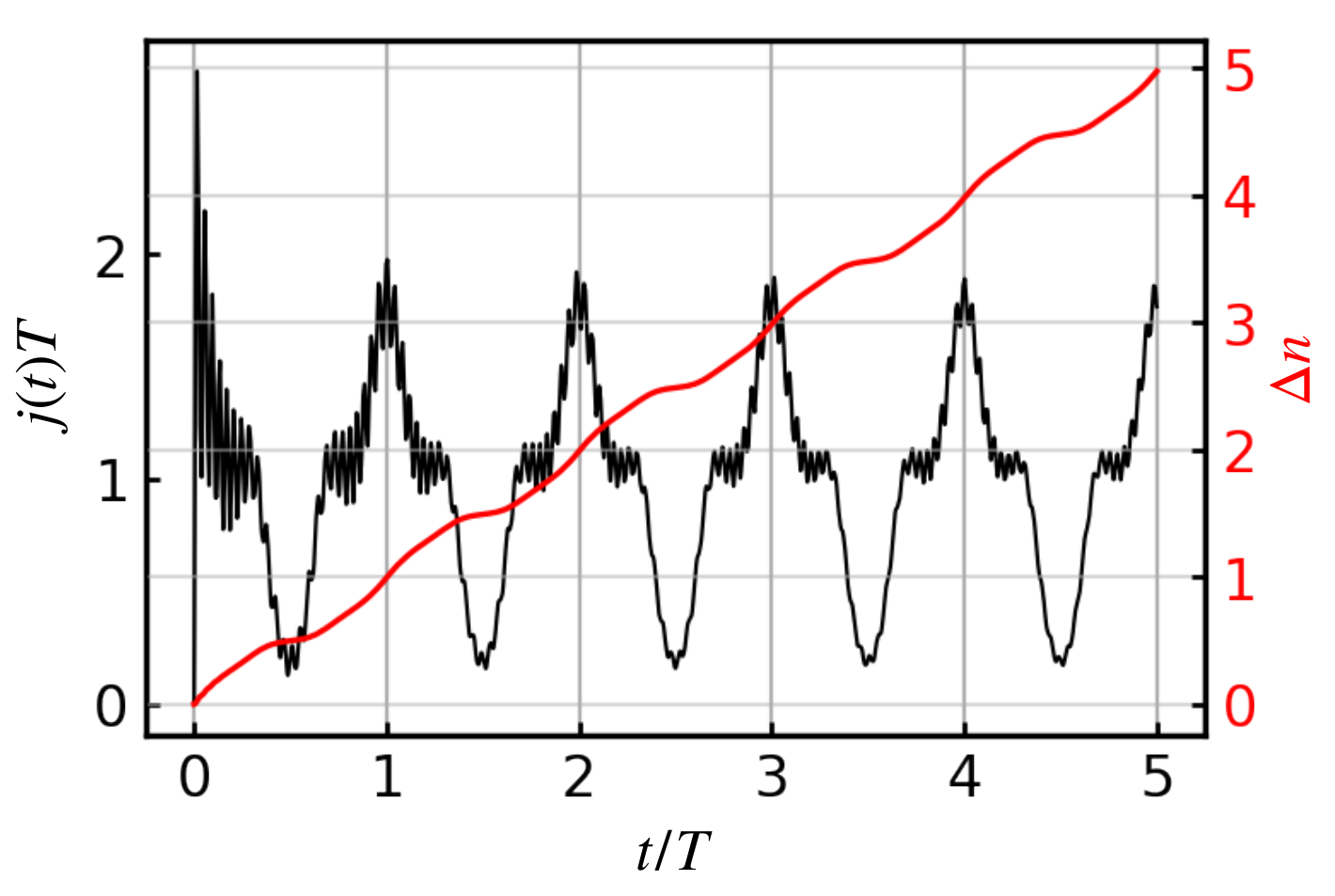}
    \caption{ Time evolution of the charge current $j(t)$ and the pumped charge $\Delta n$ for the pump scheme given in Fig. 4(b). }
\end{figure}

Turning to the pumping properties, we consider the generalized Creutz ladder with the lower band filled by spinless fermions. For simplicity, we set $t_d/t_v=0.5+0.3\cos(\omega t)$ and $\phi/\pi=-0.2\sin(\omega t)$, as labeled by the dashed curve in Fig. 4(b). As long as the trajectory in parameter space encircles the gap closing point, which is also the singular point for the Zak phase, the integrated Zak phase gives rise to an absolute value of $2\pi$, yielding a Chern number $C=1$. In time evolution, the pumped charge can be calculated by integrating the instantaneous charge current as $\Delta n(t)=\int_0^tj(t')dt'$. The charge current is defined by
\begin{equation}
	j(t)=\frac{1}{2\pi}\int_\text{1BZ}\langle\psi_k(t)|\partial\mathcal H_k(t)/\partial(\hbar k)|\psi_k(t)\rangle dk,
\end{equation}
where $\partial\mathcal H_k(t)/\partial(\hbar k)$ is the velocity operator. $|\psi_k(t)\rangle$ represents the evolved state obtained by following the time-dependent Schr\"odinger equation with an initial condition $|\psi_k(0)\rangle=|u_k(0)\rangle$, where $|u_k(0)\rangle$ is the lower band eigenstate of $\mathcal{H}_k$ at $t=0$. Results with a small driving frequency $\hbar\omega/t_v=0.05$ are shown in Fig. 6, where instantaneous current and pumped charge are plotted by black and red lines. Obviously, quantized charge pump appears at complete modulation cycles as expected.

\subsection{quantum pump by phase modulation and imbalance control}
\begin{figure}[tbp]
	\includegraphics[width=8.5cm, height=8.2cm]{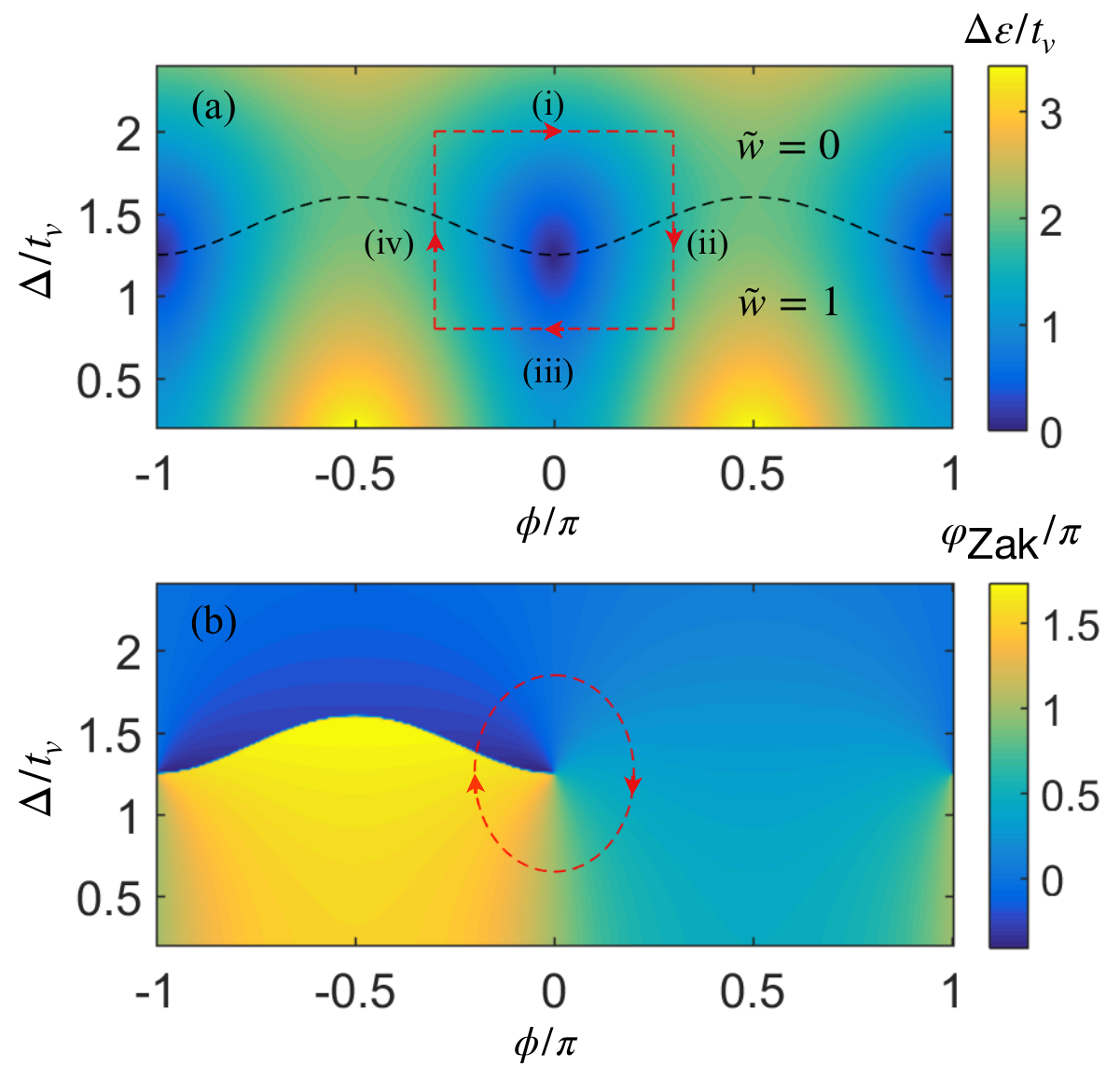}
    \caption{ Parameter space spanned by Peierls phase $\phi$ and inter-leg imbalance $\Delta$. (a) Value of band gap $\Delta\varepsilon$ with respect to $\phi$ and $\Delta$. The black dashed line locates the boundary between $\tilde w=0$ and $1$. The four paths labeled by (i)-(iv) are further demonstrated in Fig. 8. (b) Zak phase $\varphi_\text{Zak}$ as a function of $\phi$ and $\Delta$. The closed path represent a typical Thouless pumping scheme demonstrated in Fig. 9.}
\end{figure}
\begin{figure}[tbp]
	\includegraphics[width=8.5cm, height=7.4cm]{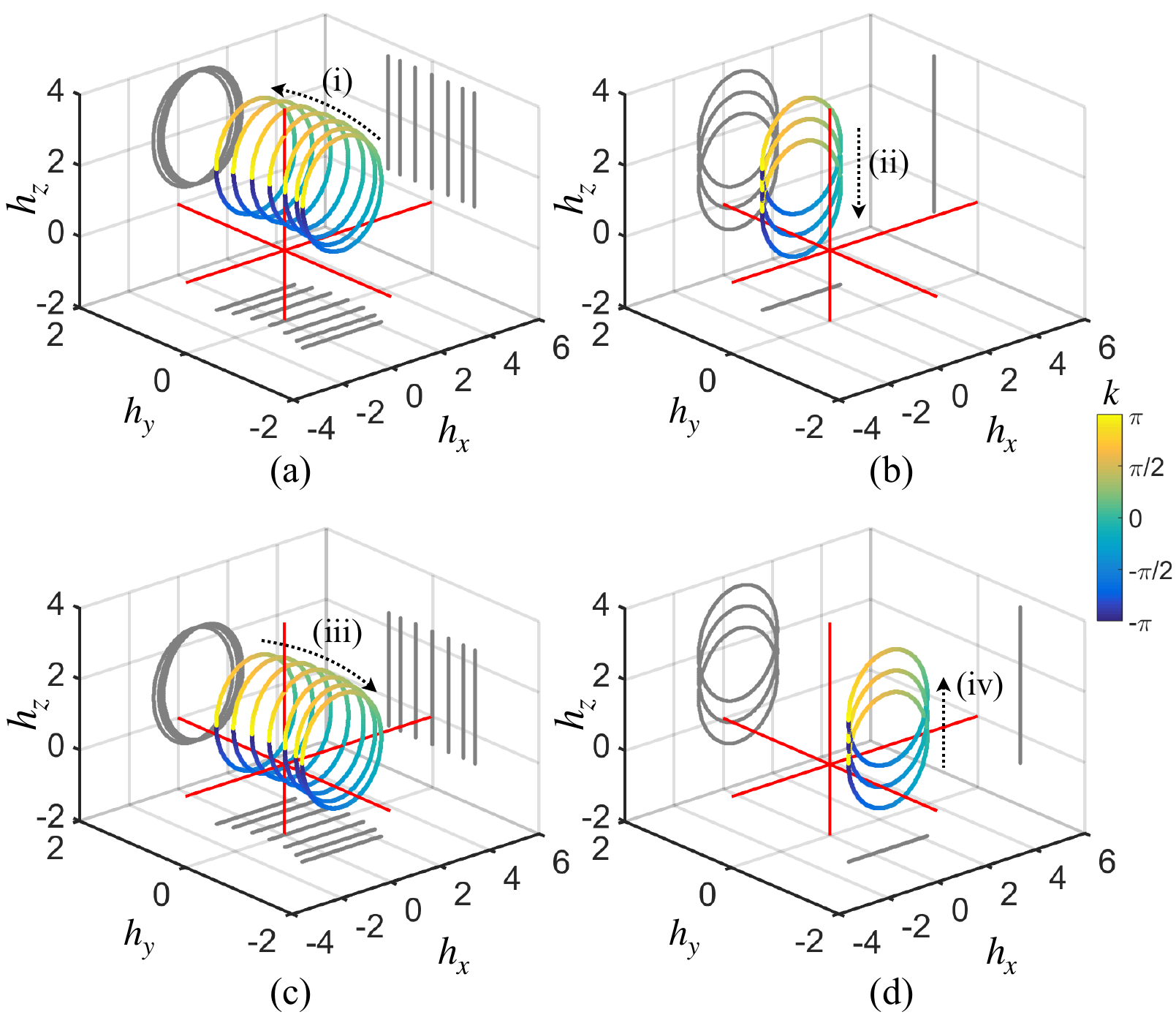}
    \caption{ Variation of the $\bm{h}$-circle following the four paths (i)-(iv) in Fig. 7. The arrows indicate the direction of changes. Projections on the bottom, rear and right side are shown. }
\end{figure}
As a second proposal, we introduce the inter-leg imbalance $\Delta$ for the system to cross the boundary between $\tilde w=1$ and $0$. The Hamiltonian is determined by
\begin{eqnarray}
	\nonumber\\
	h_x &=& t_v\cos\phi+2t_d\cos k,\nonumber\\
	h_y &=& t_v\sin\phi,\\
	h_z &=& 2t_h\sin\theta\sin k+\Delta. \nonumber
\end{eqnarray}
We have also taken $t_1=t_2=t_d$. To ensure the appearance of $\tilde w=1$ region,  we have to set $t_d/t_v>|\cos\phi|/2$. The $\bm h$-circle encompasses the $h_y$-axis ($\tilde w=1$) for $2t_h|\sin\theta|\sqrt{1-t_v^2\cos^2\phi/(4t_d^2)}>\Delta$. The band gap and Zak phase as a function of $\phi$ and $\Delta$ are shown in Fig. 7. Parameters are $t_h/t_v=t_d/t_v=0.8$ and $\theta=\pi/2$. The position of $\bm h$-circle changes with the variation of $\phi$ and $\Delta$. In Fig. 7(a), the four routes labeled by (i)-(iv) correspond to the trajectories of $\bm h$-circle demonstrated in Fig. 8(a)-(d). In this case, the imbalance provides the opportunity to lower and lift the $\bm h$-circle to cross the $h_y$-axis, instead of changing its size as in the previous case. Again, the $\bm h$-circle covers the origin once for a complete modulation cycle. We note that the evolution pattern of the $\bm h$-circle here is topologically equivalent to the mode shown in Fig. 5. Consequently, such a modulation gives rise to $C=1$.

\begin{figure}[tbp]
	\includegraphics[width=8.5cm, height=5.8cm]{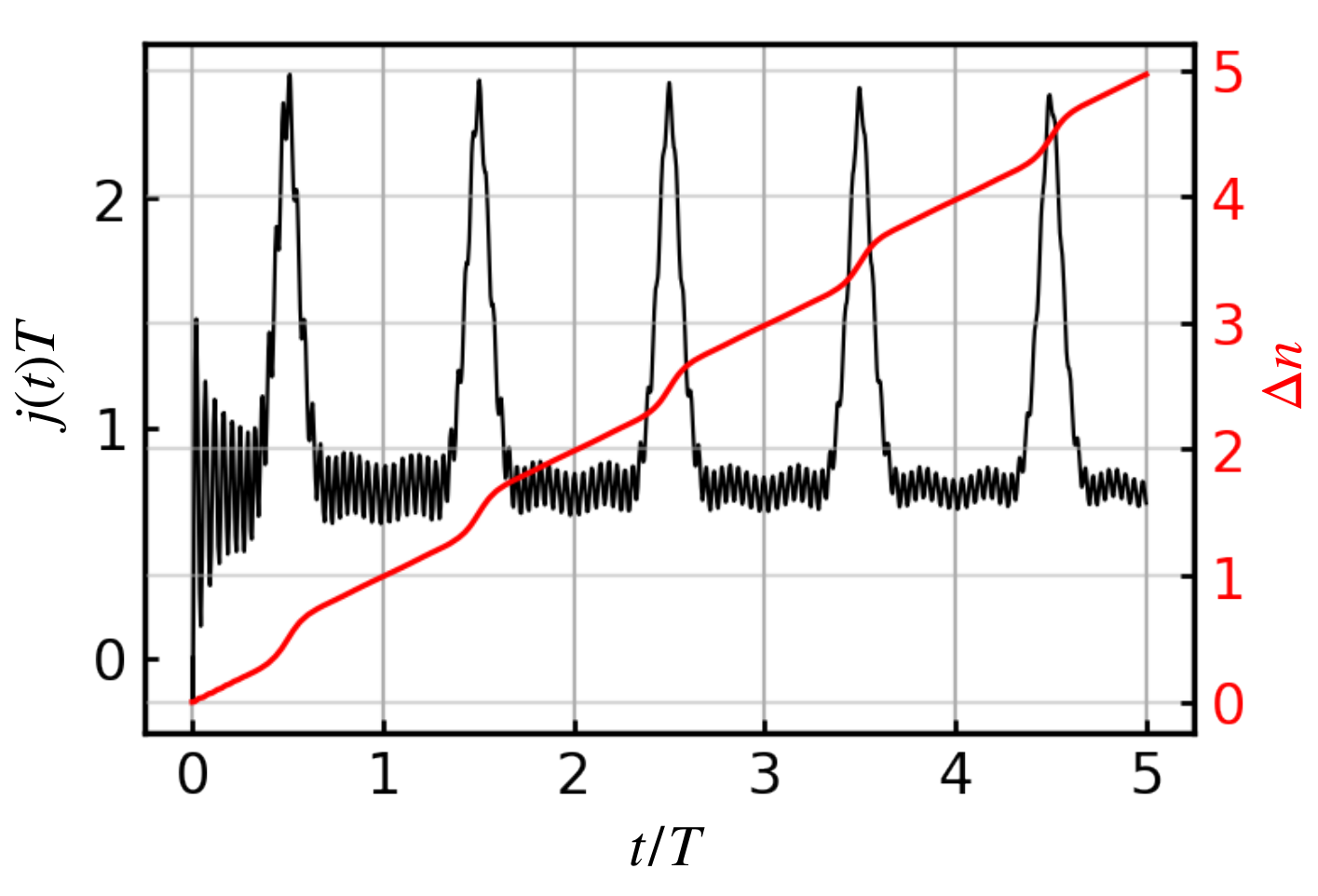}
    \caption{ Time evolution of the charge current $j(t)$ and the pumped charge $\Delta n$ for quantum pump scheme by tuning the imprinted phase and inter-leg imbalance. }
\end{figure}

To verify the quantization of charge pump, we consider the variation of parameters following the route shown in Fig. 7(b). Specifically, the modulation is given by $\Delta/t_v=2t_h|\sin\theta|\sqrt{1-t_v^2/(4t_d^2)}/t_v+0.6\cos(\omega t)$ and $\phi/\pi=0.2\sin(\omega t)$ with a period $T=2\pi/\omega$ ($\hbar\omega/t_v=0.05$). For the imbalanced ladder system with the lower band fully filled by spinless fermions, the evolution of instantaneous current $j(t)$ and pumped charge $\Delta n(t)$ over several pumping cycles are depicted in Fig. 9. Within a pumping cycle, the pumped charge depends on the specific choice of modulation, but at multiple of complete cycles, we still observe topological charge transport.

\subsection{quantum pump by hopping adjustment and imbalance control}
Let us now consider the topological charge pump induced by modulating the diagonal hopping and the inter-leg imbalance. In this case, we parameterize the diagonal hopping as $t_1=2t_d\cos^2(\gamma/2)$, $t_1=2t_d\sin^2(\gamma/2)$, and hold the Peierls phase fixed as $\phi=0$. Now the $\bm h$-circle lies in a plane which intersects with the $h_x$-$h_y$ plane at an angle $\delta$ determined by $\tan\delta=t_h\sin\theta/(t_d\cos\gamma)$. For simplicity, we have $\delta=\gamma$ by setting $t_h=t_d$ and $\theta=\gamma$. Equations (\ref{hxyz}) then reduce to
\begin{eqnarray}
	\nonumber\\
	h_x &=& t_v+2t_d\cos k,\nonumber\\
	h_y &=& 2t_d\cos\gamma\sin k,\\
	h_z &=& 2t_d\sin\gamma\sin k+\Delta. \nonumber
\end{eqnarray}
In addition, we take $t_d>t_v/2$ to ensure the appearance of non-trivial topology. The position of $\bm h$-circle can be controlled by tuning the values of $\gamma$ and $\Delta$. For $-1/2<\gamma/\pi<1/2$ [see the demonstration in Fig. 11(a)], the vector $(h_x,h_y,h_z)$ counter-clockwisely winds around the $h_z$-axis as $k$ varies across the BZ, giving rise to $\tilde w=-1$. For $\gamma/\pi<-1/2$ or $\gamma/\pi>1/2$, the winding turns to a clockwise one, indicating $\tilde w=1$. Figure 10 depicts the values of band gap and Zak phase with respect to $\gamma$ and $\Delta$. We observe gap closing points at $\gamma=\pi/2+n\pi$ ($n\in \mathcal Z$), $\Delta=\pm2t_h\sqrt{1-t_v^2/(4t_d^2)}$ and $\gamma=n\pi$, $\Delta=0$. The former locates on the boundary between $\tilde w=1$ and $-1$ (vertical dashed lines), indicating the case where the $\bm h$-circle intersects the origin, whereas the latter roots from the fact that $E_+(k=0)=E_-(k=\pm\pi)=t_v$.

\begin{figure}[tbp]
	\includegraphics[width=8.2cm, height=8.3cm]{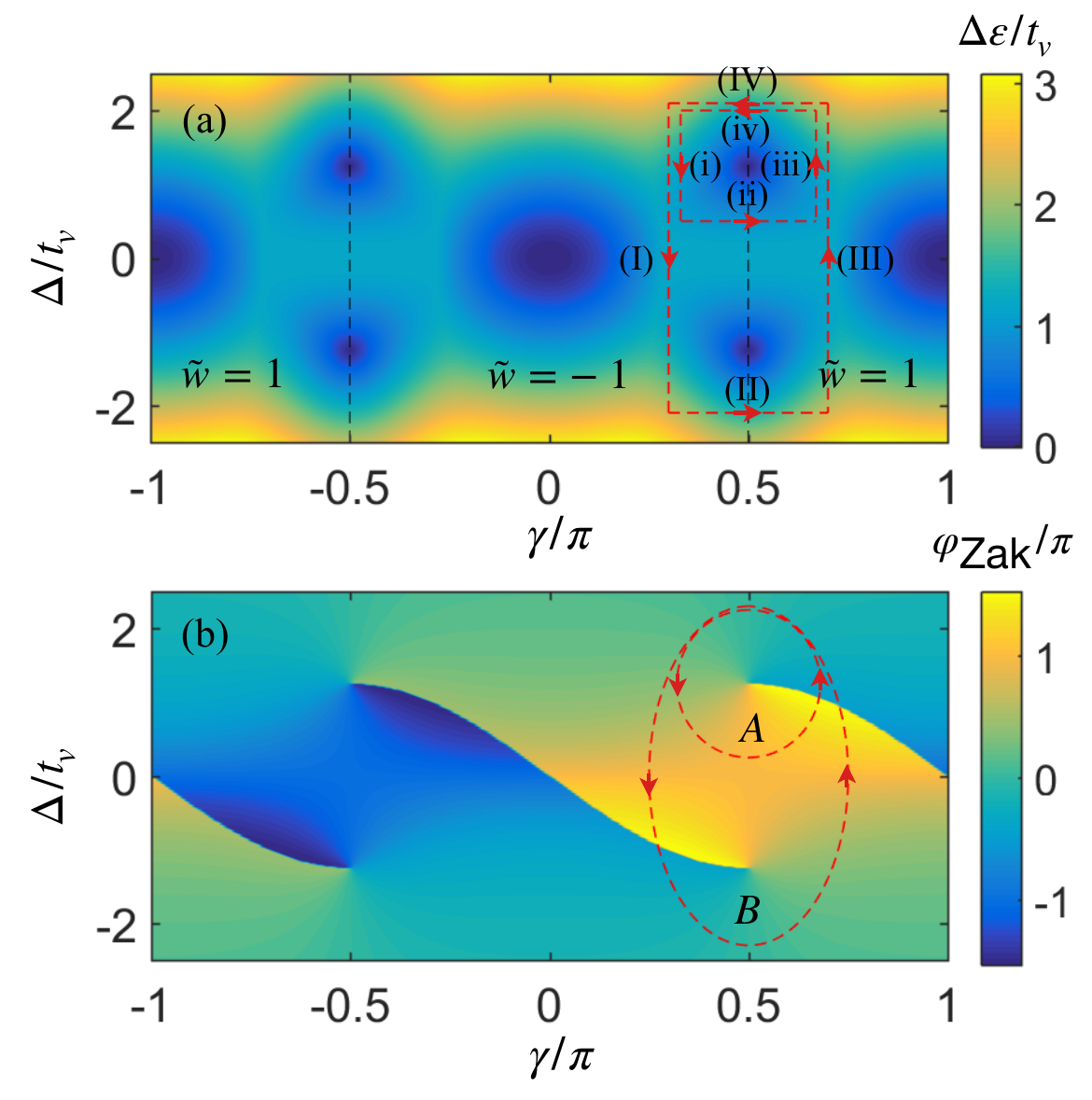}
    \caption{ Parameter space spanned by $\gamma$ and $\Delta$. (a) Value of the band gap $\Delta\varepsilon$. The black dashed line locates the boundary between $\tilde w=0$ and $1$. The paths labeled by (i)-(iv) and (I)-(IV) are further demonstrated in Fig. 11 and 12, respectively. (b) Zak phase $\varphi_\text{Zak}$ as a function of $\gamma$ and $\Delta$. The closed paths labeled by $A$ and $B$ represent two modulation schemes with different pumped charges as shown in Fig. 14. }
\end{figure}

We consider two closed paths in parameter space as labeled by (i)-(iv) and (I)-(IV) (red dashed lines) in Fig. 10(a). Following (i)-(iv), \textit{i.e.}, $(\gamma/\pi,\Delta/t_v)=(0.33,2)$ $\to$ $(0.33,0.5)$ $\to$ $(0.67,0.5)$ $\to$ $(0.67,2)$ $\to$ $(0.33,2)$, the $\bm h$-circle is lowered, rotated, lifted and rotated again to recover its original position. Such a variation is demonstrated in Fig. 11. The $\bm h$-circles shown in Fig. 11(a) and (c) refers to distinct topological features characterized by $\tilde w=-1$ and $\tilde w=1$. In Fig. 11(b) and (d), as $\gamma$ varies from $0.33\pi$ to $0.67\pi$ and vice versa, the $\bm h$-circle correspondingly rotates around the line $(h_y,h_z) = (0,\Delta)$ by an angle of $\Delta\gamma=0.34\pi$. We note that the $\bm h$-circle encompasses the origin once during process (ii), indicating that the Chern number equals $1$. This can be verified by integrating the Zak phase as shown in Fig. 10(b).

\begin{figure}[tbp]
	\includegraphics[width=8.5cm, height=8.1cm]{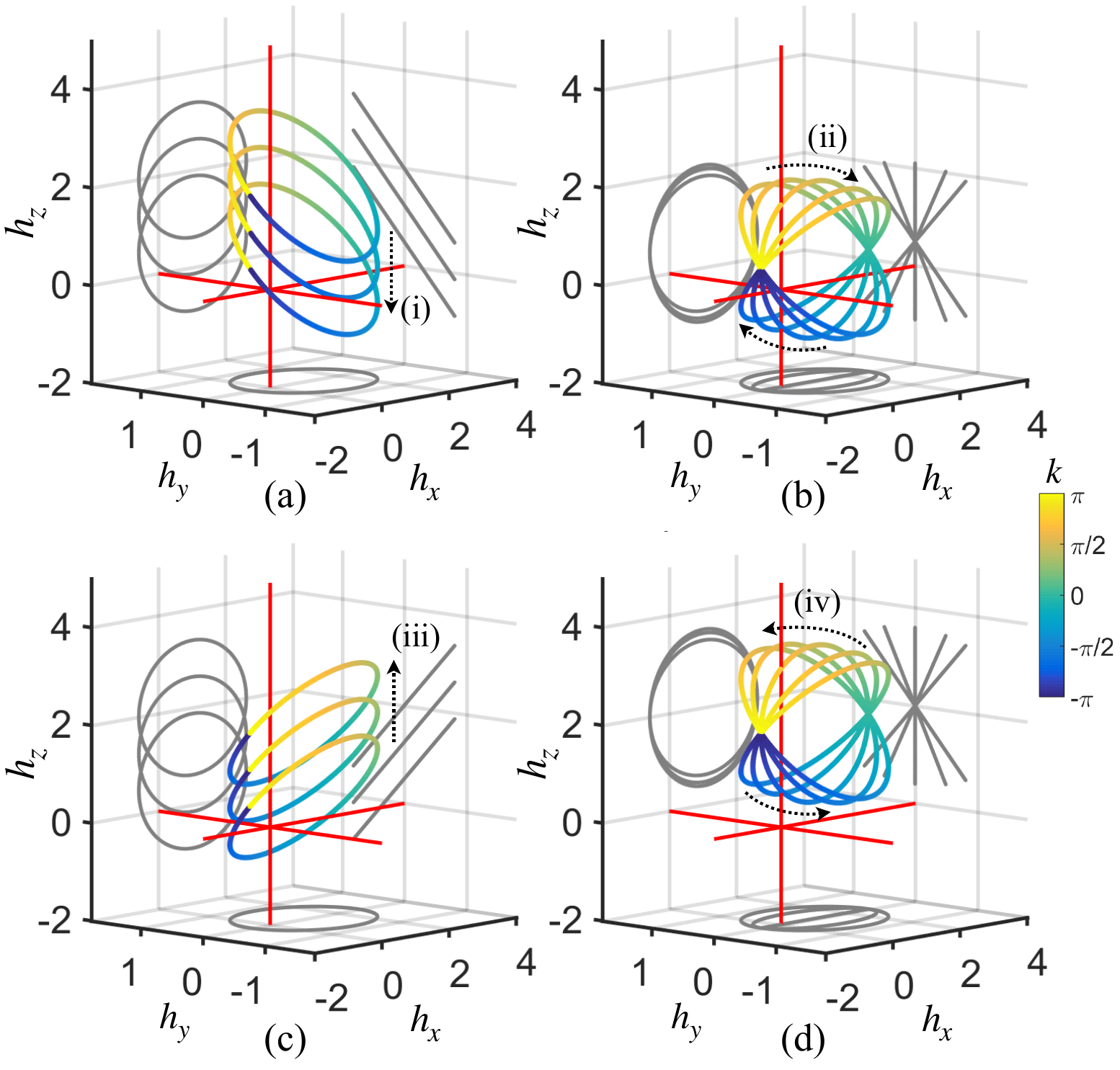}
    \caption{ Variation of the $\bm{h}$-circle following four paths (i)-(iv) in Fig. 10. The arrows indicate the direction of changes. Projections on the bottom, rear and right side are shown.  }
\end{figure}
\begin{figure}[tbp]
	\includegraphics[width=8.5cm, height=8.2cm]{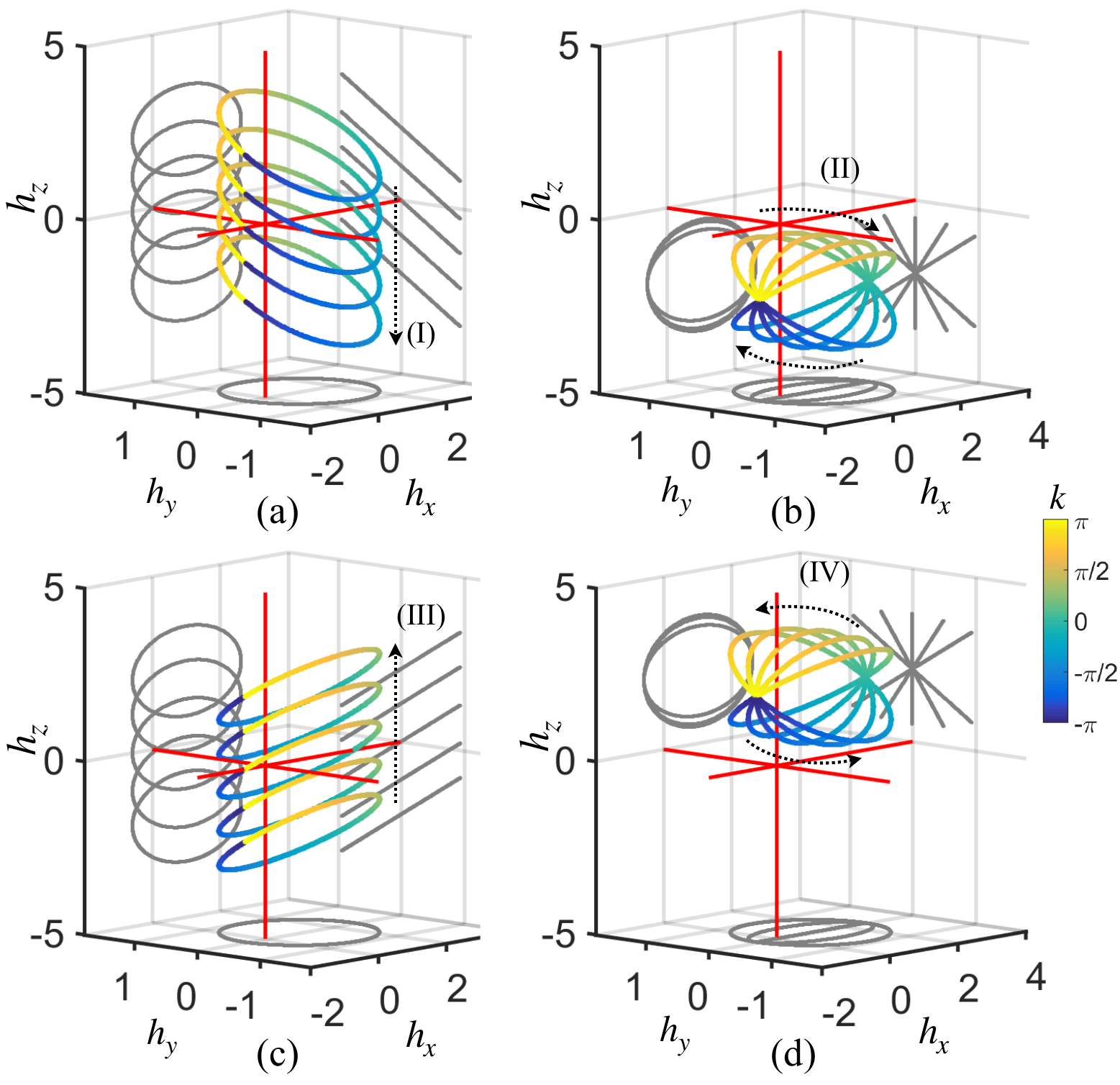}
    \caption{ Same as Fig. 11 but following (I)-(IV) in Fig. 10. }
\end{figure}

For the paths (I)-(IV), \textit{i.e.}, $(\gamma/\pi,\Delta/t_v)=(0.3,2.1)$ $\to$ $(0.3,-2.1)$ $\to$ $(0.7,-2.1)$ $\to$ $(0.7,2.1)$ $\to$ $(0.3,2.1)$, the trajectory of $\bm h$-circle is demonstrated in Fig. 12. The $\bm h$-circle experiences similar four processes as in the previous case. However, the $\bm h$-circle encompasses the origin twice during (I) and (III), as shown in Fig. 12(a) and (c), this implies that the second closed path manifests distinct topological feature characterized by Chern number $C = 2$. We mention that a closed path which connect the regimes of $\tilde w=-1$ and $\tilde w=1$ can but not necessarily give rise to $C=2$. To illustrate their topological features graphically, we hold the $\bm h$-circle fixed in Bloch space and regard the motion as a relative variation of the origin (which is a topological defect). Consequently, the two paths (i)-(iv) and (I)-(IV) correspond to very different trajectories of the origin, as shown in Fig. 13(a) and (b). Obviously, the two linking patterns are topologically distinct. Their linking numbers are $1$ and $2$, respectively, in accordance with $C=1$ and $2$. Therefore, figure 11 and 12 exhibit two typical shifting configurations of the $\bm h$-circle which is directly related to the corresponding Chern number. This provides a practical way in constructing Thouless pumping schemes.

\begin{figure}[tbp]
	\includegraphics[width=8.5cm, height=4.1cm]{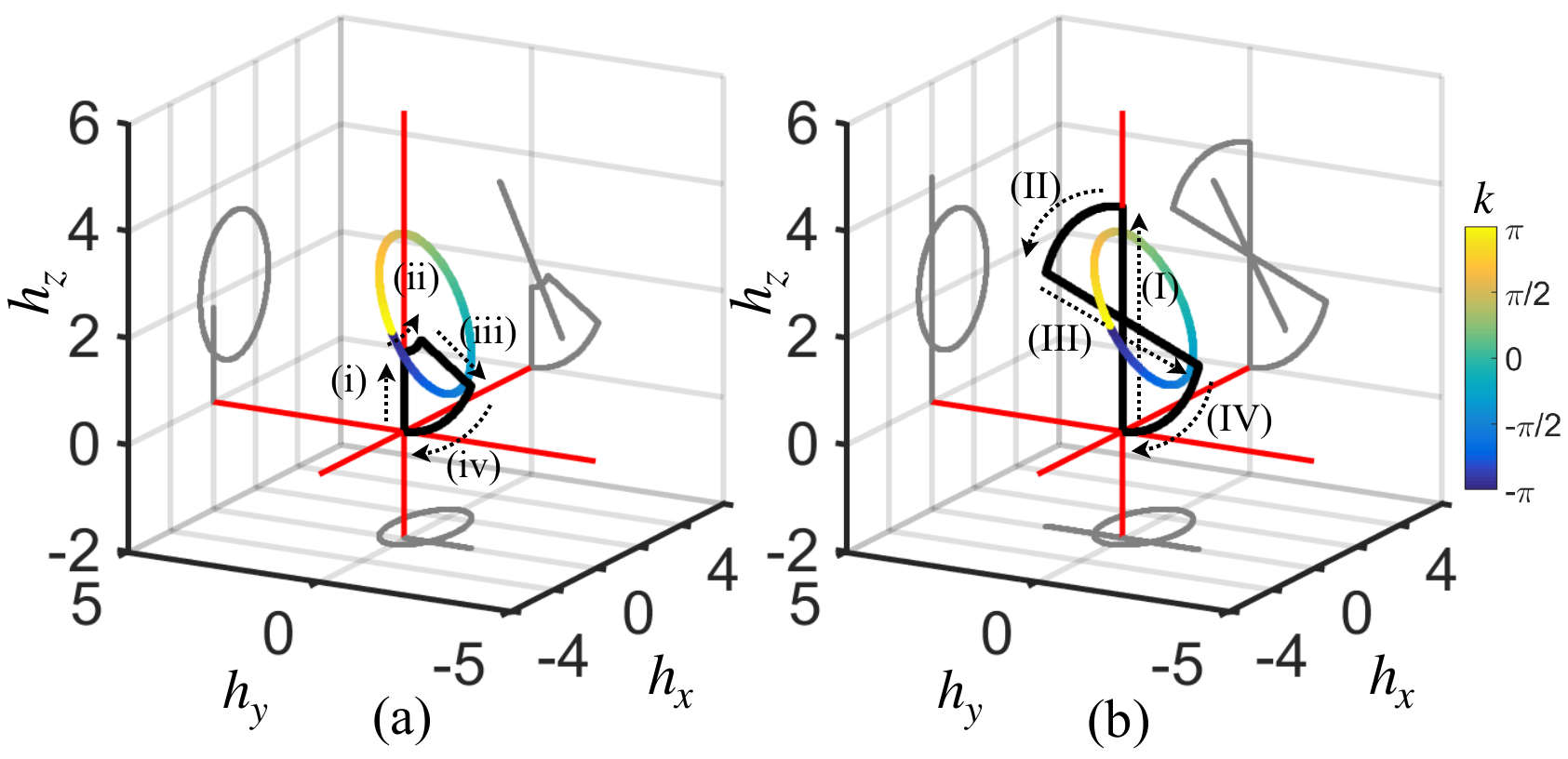}
    \caption{ Linking pattern for paths (a) (i)-(iv) and (b) (I)-(IV). The black lines plot the relative trajectory of the origin with respect to the $\bm{h}$-circle at $t=0$. Such a trajectory link with the $\bm{h}$-circle once in (a) and twice in (b).}
\end{figure}
\begin{figure}[tbp]
	\includegraphics[width=8.5cm, height=6.5cm]{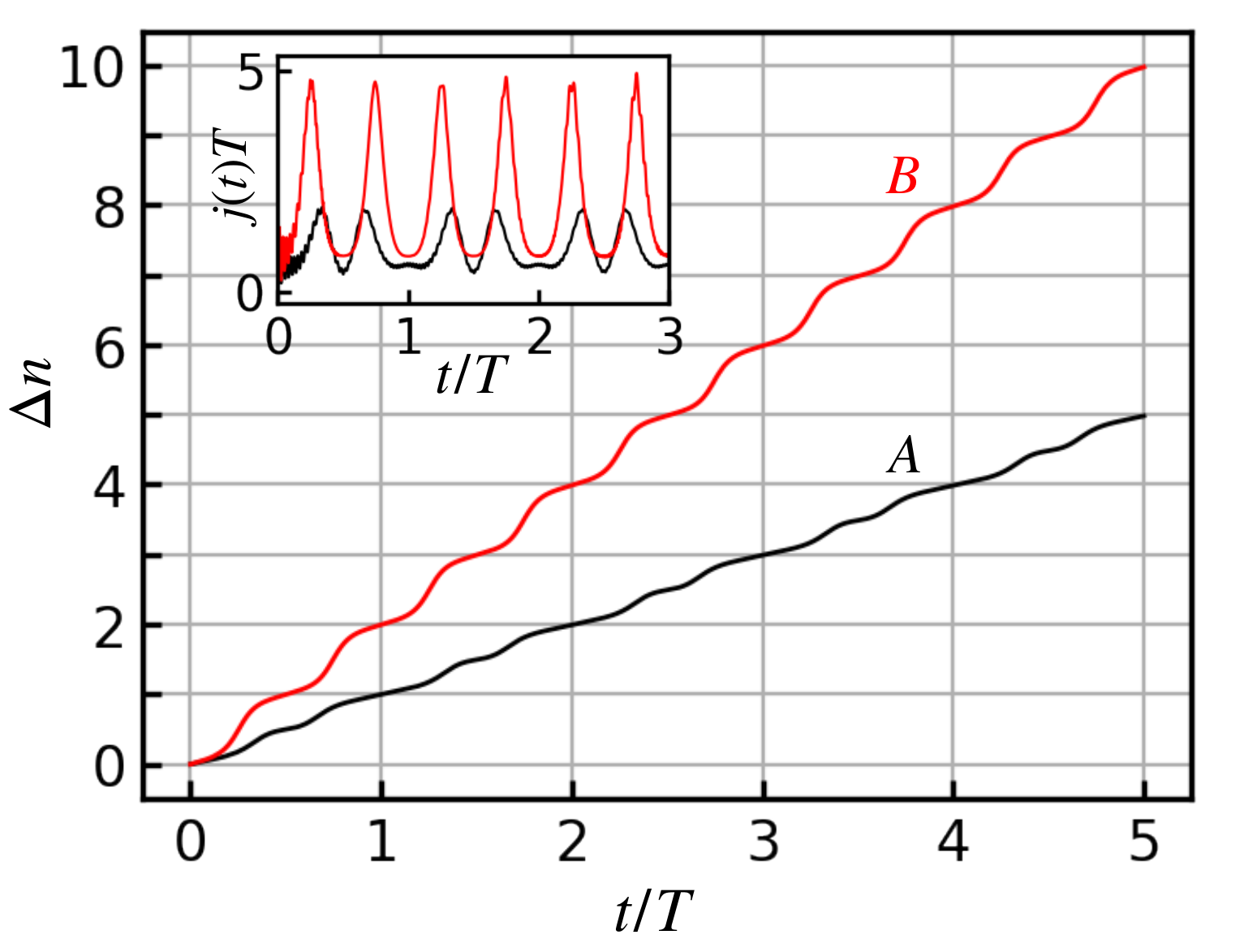}
    \caption{ Pumped charge $\Delta n$ up to time $t$ for quantum pump scheme $A$ and $B$ giving rise to $C=1$ and $C=2$, respectively. The inset shows the evolution of charge current over three periods. }
\end{figure}

Concerning the dynamical evolution, with parameters following a closed path, the pumped charge per modulation cycle is quantized to the Chern number. As a consequence of band topology, such a closed path can be deformed continuously without crossing topological defects. Analogously to the previous subsections, here we consider two cases of time-dependent evolution with parameters modulated by $A$: $\gamma/\pi=0.5-0.18\sin(\omega t)$, $\Delta/t_v=2t_h\sqrt{1-t_v^2/(4t_d^2)}/t_v+\cos(\omega t)$ and $B$: $\gamma/\pi=0.5-0.25\sin(\omega t)$, $\Delta/t_v=2.3\cos(\omega t)$. The adiabatic condition is satisfied by setting a small driving frequency $\hbar\omega/t_v=0.05$ The two paths in parameter space are depicted by dashed curves in Fig. 10(b). We set the initial state as the the lower band eigenstate of $\mathcal{H}_k(t=0)$ and carry out the simulation of the time-dependent Schr\"odinger equation. The pumped charge $\Delta n(t)$ with respect to evolving time is shown in Fig. 14. The two curves correspond to pumping $A$ and $B$, respectively. The instantaneous charge currents of the first 3 cycles are shown in the inset. As expected, the pumped charge over a modulation cycle is identical to the Chern number of the associated band. In addition, a closed path only encircling the gap closing point $(\gamma/\pi,\Delta/t_v)=(0,0)$ gives rise to $C=0$. We have verified that the time evolution following such a path yields $\Delta n=0$ at multiple of complete cycles.

\section{summary}\label{summary}\label{sec5}
Thouless pumping in crystals has emerged as a powerful tool for the manipulation of matter waves, with potential applications in several areas of physics such as robust quantum state transfer and quantum information processing \cite{dlaska2017robust, li2018unconventional}. Upon realizing the Thouless pumping, the pumped charge for a 1D or quasi-1D lattice system is quantized to the topological invariant Chern number of the associated band. In this work, we have presented a graphical representation for the topological character of a static or a modulated system, which can be easily applied to various quantum models including the Rice-Mele model, the Haldane model, as well as the generalized Creutz model we have studied. The layout of the polarization vector $(h_x,h_y,h_z)$ in Bloch space gives a quantized winding number $\tilde w$ even in the absence of chiral symmetry. By modulating parameters to connect the regimes with different value of $\tilde w$, the trajectory for the endpoints of $(h_x,h_y,h_z)$ exhibits typical patterns. These patterns directly relate to the linking configuration of two ``circles" formed by the endpoints of $(h_x,h_y,h_z)$ and the relative path of the defect (the origin). One can easily obtain the Chern number by counting the corresponding linking number. Such a graphical description facilitates the construction of quantized pumping schemes.

As a demonstration, we have studied a generalized Creutz ladder with additional Peierls phase, inter-leg imbalance and isotropic diagonal hopping. To realize Thouless pumping, three different modulation strategies are implemented. We showed how the Chern number is related to the variation of $(h_x,h_y,h_z)$ and to the linking number. As a result, pumped charge per modulation cycle is quantized with either $C=1$ or $2$. The generalized Creutz model is amenable to simulations with cold atoms in the spirit of synthetic dimension \cite{celi2014synthetic, mancini2015observation, stuhl2015visualizing, barbarino2016synthetic}. Further extension of this study to more exotic models and to higher-order topology could be envestigated.

\section{Acknowledgements}
This work is supported by the National Natural Science Foundation of China under Grants No. 12304179. Y.-K. L. acknowledges the foundation from the Fundamental Research Program of Shanxi Province under Grants No. 202103021224259.
\bibliography{TP}

\end{document}